# Strong and Highly Switchable Soft Sticky Adhesives

Qianfeng Yin[1], Yilmaz Arin Manav[2], Yichen Wan[1], Benyamin Davaji[2,*], and Ruobing Bai[1,*]

[1]Department of Mechanical and Industrial Engineering, College of Engineering, Northeastern University, Boston, MA, 02115, USA.

[2]Department of Electrical and Computer Engineering, College of Engineering, Northeastern University, Boston, MA, 02115, USA.

[*]Corresponding author: Benyamin Davaji: b.davaji@northeastern.edu; Ruobing Bai: ru.bai@northeastern.edu

**Author Contributions:** Q.Y. and R.B. designed the research; Q.Y., Y.A.M, and Y.W performed the research; Q.Y., Y.A.M analyzed the data; Q.Y. and R.B. wrote the first draft of the paper; all authors contributed to the writing of the final manuscript.

**Competing Interest Statement:** A provisional patent application, titled "THERMO-SWITCHABLE PRESSURE-SENSITIVE ADHESIVES", has been filed with the U.S. Patent and Trademark Office, based on the current work.

**Abstract**

Many biological systems can form strong adhesion to various materials with complex shapes. The adhesion is further switchable between strongly adhering and completely non-adhering in a simple and fast manner. By contrast, no engineering system has yet achieved the same robust adherence and switching. This limitation severely hinders the advancement of several emerging technologies including biomimetic robots, assembly-based manufacturing, precision medicine, wearable and implantable devices, as well as on-demand material dismantling and recycling for sustainability. Here we present a design approach for strong and highly switchable adhesion by synergizing the surface stickiness, bulk energy dissipation, and stimuli-responsive polymer chains in a thermo-switchable soft sticky adhesive. The adhesive has a high adhesion strength of about 80 kPa with diverse materials at room temperature. The adhesion is highly switchable to near-vanishing (about 0.6 kPa) at an elevated temperature due to the thermo-responsive surface polymer chain retraction. This adhesion switching is reversible and repeatable for many cycles, enabling selective pick-and-release of objects with various materials, shapes, sizes, and weights. The switching time is around 10 s with an adhesive layer of 1 mm, governed by thermal conduction through the adhesive, faster than or comparable to most state-of-the-art methods. The adhesive is self-healing, and can be recycled, dried, stored, reswollen, and reused with nearly intact adhesion and switching properties. The synergistic design combining strong adhesion and stimuli-responsive switching can be potentially extended to various polymer systems, and further enhanced by optimized surface architectures.

**Introduction**

Many biological systems can form strong adhesion to external objects made of various materials with complex shapes. The adhesion is further switchable between strongly adhering and completely non-adhering. Examples include the fibrillar-structured attachment pads of beetles, flies, spiders, and geckos (1). Developing strong and highly switchable adhesion in human-engineered systems is attractive in advancing a broad range of emerging technologies, such as biomimetic climbing and flying robots in complex environments (2, 3), additive manufacturing via rapid assembly of functional voxels (4, 5), precision in-body wound dressing and drug delivery (6, 7), and wearable and implantable biomedical devices with multifunctional human-machine interfaces (8, 9).

An engineering system can achieve switchable adhesion in multiple ways, including pneumatic suction (10, 11), electrostatic force (2, 12), modulus regulation (13, 14), mechanical interlocking (15-17), and gecko-inspired micro-pillar surfaces (18, 19). Compared to these existing methods that have been actively explored in recent years, enabling switchable adhesion in a soft sticky polymer network, commonly called a *pressure-sensitive adhesive* (PSA) in commercial applications (20), has just started showing its unique advantages (21-23). A switchable soft sticky adhesive can adhere to small and non-flat objects made of various materials, which is not achievable by most other systems (e.g., pneumatic or electrostatic forces). For an adhesive that is sticky by default and switchable to non-sticky by an external stimulus, the system only consumes energy during the switching process, greatly improving the energy efficiency for large-scale deployment such as manufacturing via high-throughput pick-and-release. Commercial PSAs have been widely used in packaging, textile, electronics, manufacturing, and civil infrastructure (20). As a result, switchable PSAs are promising to address the emerging global challenge of material sustainability, by enabling on-demand dismantling and recycling of devices to reduce the 7 to 10 billion tons of global waste produced every year (24, 25).

While strong, irreversible adhesion has been studied for decades (26), the investigation on switchable adhesion has only attracted significant attention in recent years (see recent reviews and progress reports (24, 25, 27-29)). To date, no engineering system has yet achieved the same robust adherence *and* switching that is comparable to a biological system, suffering from drawbacks including narrow switching ranges (a strong adhesion of about 100 kPa still leaves nontrivial residual adhesion over 10 kPa when switched off), complex switching processes (switching requires multiple steps of combined stimuli and mechanical debonding), stringent working conditions (special chemistries, surface bonds, or adhering materials), and low switching speed (switching takes > $10^2$ s in most systems) (24, 25, 27-29).

To address these existing issues, here we present a design approach of synergizing surface stickiness, bulk energy dissipation, and stimuli-responsive polymer chains to enable strong and highly switchable soft sticky adhesives. To illustrate our hypothesis, we synthesize a thermo-switchable PSA-like soft sticky adhesive made of a thermo-responsive poly(N-isopropylacrylamide) (PNIPAm) network crosslinked by nanoclay. We experimentally investigate the detailed synthesis-property relationship including the effects of initiator, accelerator, nanoclay crosslinker, and water content on the adhesion and its switching. The soft sticky adhesive has an adhesion strength of about 80 kPa with diverse materials at room temperature. The adhesion is highly switchable in a facile way: at an elevated temperature over 32 °C, the adhesion strength drops to a near-vanishing 0.6 kPa due to the substantial thermo-induced chain retraction at the interface. The adhesion switching is reversible and repeatable for many cycles. Integrated into a customized gripper with predesigned heating zones, the adhesive enables selective pick-and-release of objects with various materials, shapes, sizes, and weights. The switching time is governed by thermal conduction through the adhesive layer, which is around 10 s with an adhesive layer of 1 mm. The adhesive is self-healing, and can be recycled, dried, stored, reswollen, and reused with nearly intact properties. The synergistic design combining strong adhesion and stimuli-responsive switching can be potentially extended to various polymer systems, and further enhanced by optimized surface architectures.

**Results**

**Design principle for strong and switchable soft sticky adhesives.** Our hypothesized design principle is two-fold. First, a PSA-like soft sticky polymer network has many long free-end dangling chains on its surface. These chains can form densely packed noncovalent (e.g., van der Waals) bonds with various materials (herein called *adherends*) without any special surface treatment, leading to strong adhesion (**Figure 1** left). During debonding, stretchable fibrils emerge from the polymer network (**Figure 1** left, **Movie S1**), bridge the crack at interface, transmit the local stress to the bulk polymer network, and induce large energy dissipation via the breaking of other sacrificial bonds in the bulk (26, 30). To switch off the adhesion, the sticky polymer network undergoes a reversible stimuli-responsive phase transition, significantly changing its bulk or surface property such as a non-sticky surface due to chain retraction (**Figure 1** right), leading to greatly reduced or vanishing adhesion.

To illustrate this design principle, we synthesize a thermo-switchable sticky poly(N-isopropylacrylamide) (PNIPAm) hydrogel crosslinked by nanoclay as the adhesive. We use nanoclay as crosslinkers since they form a large number of physical (noncovalent) bonds with PNIPAm chains (31, 32), enabling a highly viscoplastic polymer network with large bulk energy dissipation during debonding. We finely tune the amount of initiator, accelerator, crosslinker, and water content in the synthesis to ensure a strong adhesion with diverse materials at room temperature. In switching, thermal stimulus is used since heat can be easily converted from various other stimuli including electricity, magnetic field, and light. At temperature above 32 °C, the rubbery PNIPAm network transforms to a condensed glassy-like network (see the process of this phase transition in **Figure S1** and **Movie S2**) (33, 34), inducing significant polymer chain retraction at the interface. This is evidenced by a completely non-sticky PNIPAm network at high temperature, with no fibril forming during the debonding from an adherend, indicating a near-vanishing adhesion after switching (**Figure 1** right, **Movie S1**).

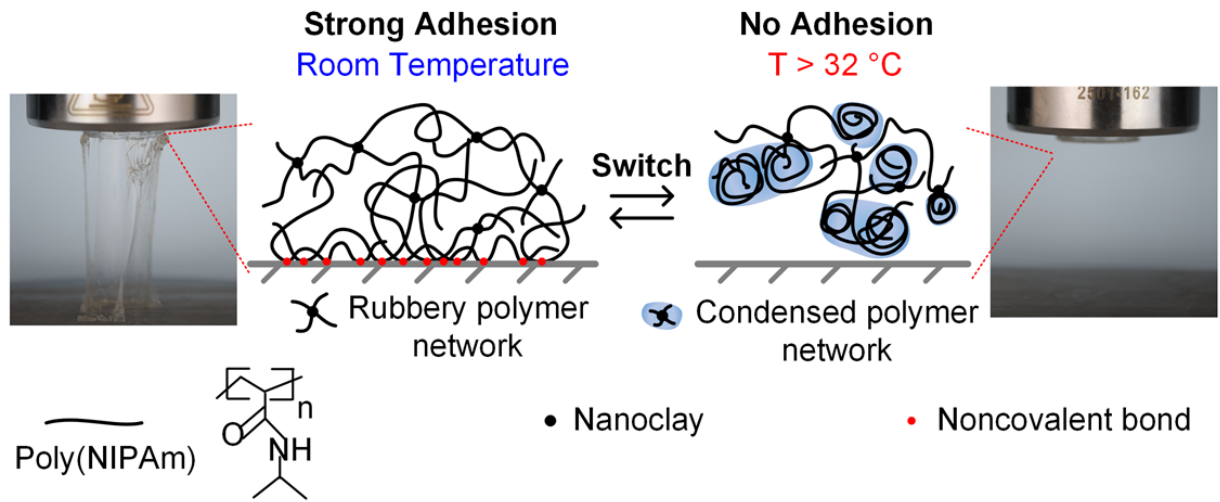

**Figure 1.** A strong and thermo-switchable soft sticky PNIPAm adhesive based on the hypothesized design principle. Left: at room temperature, the adhesive is sticky at the surface and highly viscoplastic in the bulk, providing strong adhesion, as indicated by the stretchable fibrils emerging from the polymer network during debonding from an adherend. Right: at an elevated temperature ($T >$ 32 °C), the rubbery PNIPAm network transforms to a condensed glassy-like network, inducing significant chain retraction at the interface, leading to near-vanishing adhesion with no fibril. The complete debonding processes at the two temperatures are shown in **Movie S1**.

**Synthesis-property relationship for strong adhesion.** To provide strong adhesion at room temperature, the polymer network needs many long dangling chains on its surface to enable sufficient interfacial bonds, a percolated network in its bulk to effectively transmit the stress from the interface, and additional bulk energy dissipation from the breaking of physical bonds between nanoclay and polymer chains. These processes highly depend on the synthesis parameters including the amounts of initiator, accelerator, nanoclay crosslinker, and water solvent. During synthesis, NIPAm monomers interlink to form PNIPAm chains via free-radical polymerization assisted by the initiator and accelerator, and PNIPAm chains simultaneously crosslink into a network by forming physical bonds with nanoclay. In experiment, we fix the mass of NIPAm monomers, and vary the amounts of initiator (ammonium persulfate, APS), accelerator ($N,N,N',N'$-tetramethylethylenediamine, TEMED), nanoclay, and water solvent to synthesize PNIPAm with different compositions (See **Materials and Methods** for the detailed experimental process). We characterize their adhesion strength using the probe-pull experiment (**Figure S2**) and adhesion energy using the 90-degree peeling or double-peeling experiment (**Figure S3**), both at room temperature.

We first fix all other parameters and vary the mass ratio of initiator $M_{APS}/M_{NIPAm}$. The synthesized PNIPAm always reaches the highest adhesion strength (**Figure S4**) or adhesion energy (**Figure S5**) at an intermediate $M_{APS}/M_{NIPAm}$. This intermediate value is slightly different for adhesion strength and adhesion energy, and depends on other fixed parameters. The intermediate optimal amount of initiator can be understood by the following molecular picture. When $M_{APS}/M_{NIPAm}$ is too low, during polymerization, a percolated polymer network cannot form due to insufficient initiating points, leading to weak or no adhesion. When $M_{APS}/M_{NIPAm}$ is too high, many polymer chains are initiated simultaneously during polymerization, but quickly terminated by neighboring initiators, leading to many dangling chains with short chain lengths (35). The short chains result in PNIPAm with low adhesion strength, following the classical Lake-Thomas model for fracture (36, 37), consistent with our experimental observation. This non-monotonic effect of initiator is further coupled with other synthesis parameters such as the mass ratios of nanoclay (**Figure S4A&S5A**) and accelerator (**Figure S4B&S5B**). For example, the adhesion is greatly enhanced with the increase of accelerator, which is likely due to the increased number of dangling chains in the system.

We next fix the amounts of initiator and accelerator, and investigate the effects of nanoclay (**Figure 2A & Figure S6**) and water content (**Figure 2B**). Overall, both the adhesion strength and adhesion energy increase with the mass ratio of nanoclay, $M_{clay}/M_{NIPAm}$, since more nanoclay leads to more physical bonds with polymer chains, enhancing the bulk energy dissipation during debonding. This effect of nanoclay is consistently observed in adhesion strengths with various adherends (**Figure S7**). However, at certain compositions, too much nanoclay leads to decreased adhesion strength (**Figure S6A**) or adhesion energy (**Figure S6B**). This is possibly due to the competing effects of nanoclay as crosslinkers and dissipators: when the amount of nanoclay increases, the former embrittle a network and the latter toughen a network. Such competing effects have been observed and discussed in other well-studied soft material systems such as the double-network polyacrylamide-calcium-alginate hydrogel (38, 39). Besides nanoclay, the adhesion strength increases with the mass ratio of water solvent $M_{water}/M_{gel}$ until a peak value, and subsequently decreases (**Figure 2B**). In experiment, a low water content (e.g., < 50 wt%) leads to a non-sticky surface, likely due to the collapse of free-end dangling chains on the surface. On the other hand, a too high water content (e.g., > 94 wt%) causes large swelling of the gel, leading to lower adhesion due to decreased surface density of dangling chains and a more brittle bulk material (40). Finally, with all the synthesis parameters fixed (see **Materials and Methods** for the final parameters for the rest of this paper), the adhesion strength increases with the compressive pressure applied to attach the adhesive to an adherend (**Figure S8**). This *pressure-sensitivity* again alludes to the nature of densely packed physical bonds at the interface—a high pressure during attachment promotes the contact between the dangling polymer chains and adherend.

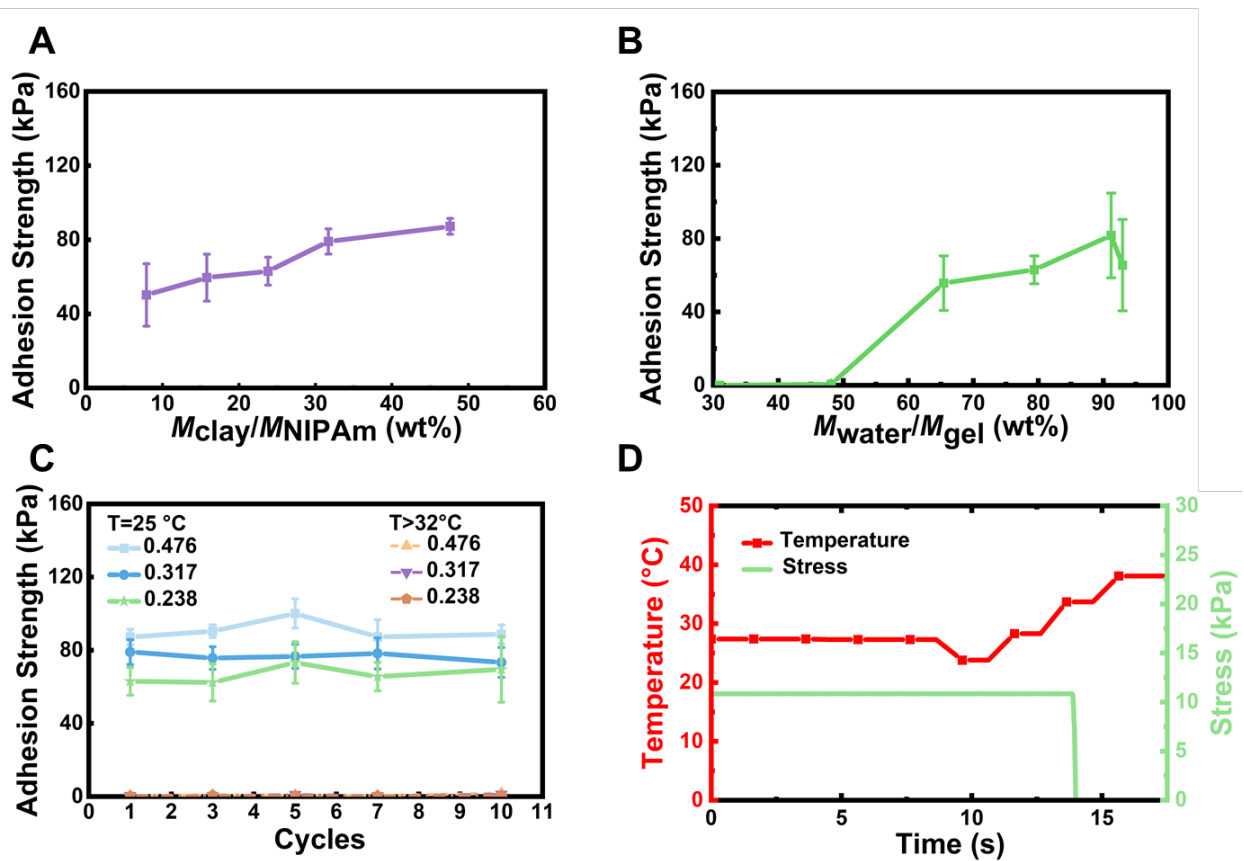

**Figure 2.** Synthesis-property relationship for adhesion strength and characterization of adhesion switching. The adherend is chosen as glass for all the data. **(A)** The adhesion strength increases with the mass ratio of nanoclay, $M_{clay}/M_{NIPAm}$. The amounts of initiator and accelerator are fixed as $M_{APS}/M_{NIPAm} = 3.818×10^{-3}$ and $M_{TEMED}/M_{NIPAm} = 72.51×10^{-3}$, respectively. **(B)** The adhesion strength increases with water content, reaches a peak, and then decreases. **(C)** The adhesive is switchable for 10 consecutive cycles, showing high adhesion strength at room temperature ($T = 25$ °C) and near-vanishing adhesion strength (about 0.6 kPa) at high temperature ($T > 32$ °C). The different values in the legend represent different mass ratios of nanoclay, $M_{clay}/M_{NIPAm} = 0.476, 0.317$, and $0.238$. **(D)** Measurement of switching time. The constant stress plateau (green curve) represents the adhesion stress generated by a dead weight being attached to the adhesive gripper, which is further lifted by the tensile tester. The temperature is measured by the *in situ* temperature sensor on the heater (red curve). The switching time (14 s) is measured as the interval from activating the heating to the moment of sudden drop of the stress. All error bars represent the standard deviation of more than 3 independently measured samples.

**Switchable adhesion for many cycles.** The synthesized adhesive is highly switchable between about 87 kPa at room temperature and nearly no adhesion (about 0.6 kPa) at temperature above 32 °C (**Figure 2C**), due to the thermo-responsive phase transition of PNIPAm that has been well studied (33). To quantitatively validate this critical temperature of 32 °C, we build a homemade gripper with the adhesive attached to its bottom end. As shown by **Figure S9** and **Movie S3** with the assist of thermal imaging, the gripper can successfully pick up an aluminum plate at room temperature, but cannot pick up an identical plate that has been preheated to precisely 32 °C. The room-temperature plate attached to the gripper is subsequently heated. When its temperature increases to slightly above 32 °C, the adhesive undergoes phase transition indicated by the opaque color, and is easily detached from the plate (**Movie S4**). This thermal switching is completely reversible for 10 switching cycles (**Figure 2C**, see **Materials and Methods** and **Figure S10** for the characterization setup). Over many cycles of switching, the adhesive will gradually lose water due to the continuous heating. Nevertheless, with a slight spray of water on its surface after each switching cycle, the lifetime of switching can be greatly extended, e.g., more than 30 cycles as shown in **Figure S11**.

To further characterize the switching time in a practical application of pick-and-release, we integrate the homemade adhesive gripper with a heater and an *in situ* temperature sensor (**Figure S12**). We attach a dead weight onto the gripper, and lift the gripper, adhesive, and weight using an Instron tensile tester (**Figure S13A**). At room temperature, the gripper can hold the weight for more than 100 s, shown by the constant stress plateau measured by the tensile tester (**Figure S13B**). By contrast, with a heating towards 40 °C, the gripper can only hold the dead weight for about 14 s (**Figure S13C** and **Figure 2D**), where the detachment occurs within 2 s once the temperature of the heater reaches 32 °C. This switching time of 14 s is consistent with the time scale of thermal diffusion throughout the thickness of the adhesive, $t \sim h^2/D_t$. Taking $h \sim 1$ mm as the adhesive thickness and $D_t \sim 10^{-7}$ m$^2$/s as the thermal diffusivity of water, we have $t \sim 10$ s. In other words, heat needs to be conducted from the heater side to the adhering side throughout the adhesive thickness to activate the polymer chain retraction. Since $t$ scales quadratically with the thickness $h$, a thinner adhesive can further reduce the switching time.

**Switchable adhesion with diverse materials.** The strong and switchable adhesion for many switching cycles apply to adherends made of diverse materials, including glass, plastics, elastomers, metals, and wood (**Figure 3 & Table S1**). For nearly all the materials investigated, the adhesion strength at room temperature is around 80 kPa in both the first and tenth cycles. The lowest adhesion strength is measured with wood, 39.34±5.22 kPa before any switching and 29.15±5.23 kPa after 10 cycles of switching. At high temperature after switching, the residual adhesion strength is mostly lower than 1 kPa, with the highest of 2.82±0.77 kPa measured with wood too. Both the lower room-temperature adhesion and higher switched residual strength for wood might be attributed to its high surface roughness: at room temperature, a rough surface reduces the effective contact area hence lowers the adhesion; when being switched at high temperature, the rough surface forms certain mechanical interlocking with the stiffer condensed PNIPAm network, giving rise to residual adhesion.

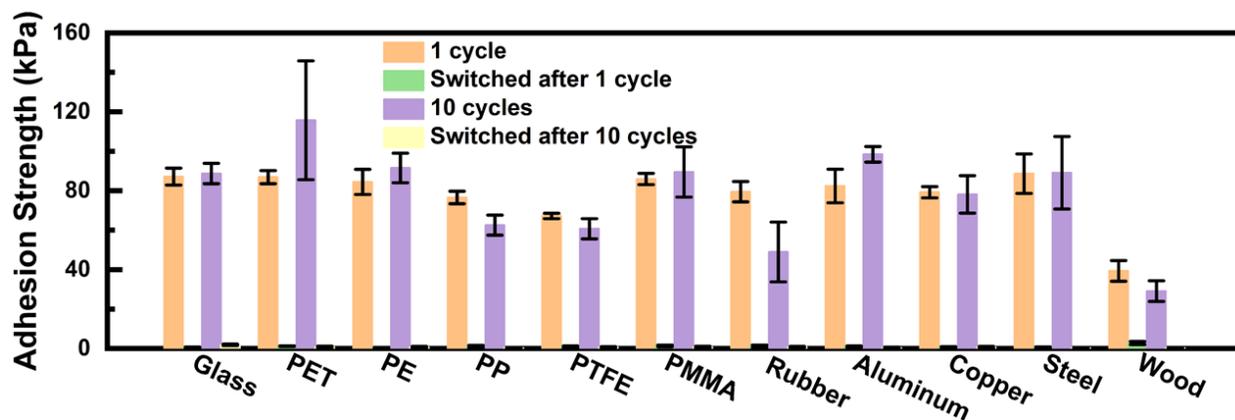

**Figure 3**. The adhesive forms strong and highly switchable adhesion with diverse materials including glass, PET (polyethylene terephthalate), PE (polyethylene), PP (polypropylene), PTFE (polytetrafluoroethylene), PMMA (poly(methyl methacrylate)), silicone rubber, aluminum, copper, steel, and wood. The adhesion at room temperature remains nearly the same after 10 cycles of switching. For all materials, the residual adhesion strengths at high temperature are around or lower than 1 kPa, which are shown by the negligible green and yellow columns. See **Table S1** for all the measured values. All error bars represent the standard deviation of more than 3 independently measured samples.

In contrast to the relatively consistent adhesion strength with various materials, the adhesion energy varies over orders of magnitude (**Figure S14**), with the highest of about 600 J/m$^2$ for PET, and the lowest of about 30 J/m$^2$ for steel. Nevertheless, these are still much higher than the adhesion energy induced by physical bonds in many other established methods such as using nanoparticles (41). To investigate this large discrepancy in adhesion energy among different materials, we conduct contact angle measurements for all the adherends, and rank them in **Figure S15** together with their corresponding adhesion strengths and adhesion energies (also see detailed measured values in **Table S2**). With the increase of contact angle, an adherend is more hydrophobic, and the adhesion strength shows a slight decrease except for the case of PE. This slight decrease of adhesion strength with the contact angle, though not significant and with certain scattering, may be attributed to the *osmocapillary* effect on adhesion that has been recently reported (42). When the adhesive gel is in contact with an adherend, their interfacial gap is filled by water due to the capillary force, and this thin layer of water is under tension due to the osmotic pressure from the gel, thus providing adhesion at the interface. This osmocapillary adhesion decreases with the hydrophobicity of the adherend, since a more hydrophobic adherend provides less capillary force to pull water out from the gel. On the other hand, a more hydrophobic adherend promotes the formation of more physical bonds between the dangling polymer chains and adherend, thus enhancing the adhesion. With the competition between these two mechanisms, we postulate that the adhesion strength is jointly influenced by the interfacial water layer and dangling chains, but the former does not effectively contribute to the adhesion energy. This is due to the poor resistance of water to shear deformation, resulting in its minimal capacity to transmit the local crack-tip stress to the bulk adhesive to elicit its energy dissipation (43). The distinct quantitative roles of the interfacial water layer and dangling chains in adhesion are being investigated in a follow-up work with additional systematic experiments, such as interfacial fracture and fatigue of gel adhesives with different water content and crosslink density.

**Self-healing and dry-stored reusable adhesive.** The adhesive shows perfect self-healing due to the reversible physical bonds between nanoclay and PNIPAm chains. After a probe-pull experiment (**Figure S16**), we re-collect the broken adhesive pieces, patch them together, reshape them using a mold, and allow them to heal for 24 hours (**Figure 4A left**). The adhesion strength after healing is nearly identical to that of the freshly made adhesive (**Figure 4A right**). Furthermore, the adhesive can be dried, grinded into small particles, stored for 10 days, re-swollen in water, patched together, and reused (**Figure 4B**). The stored adhesive almost completely recovers its adhesion strength at room temperature and excellent switching at high temperature, even after 10 switching cycles (**Figure 4C**). This complete reusability extends to adherends made of diverse materials (**Figure 4D**).

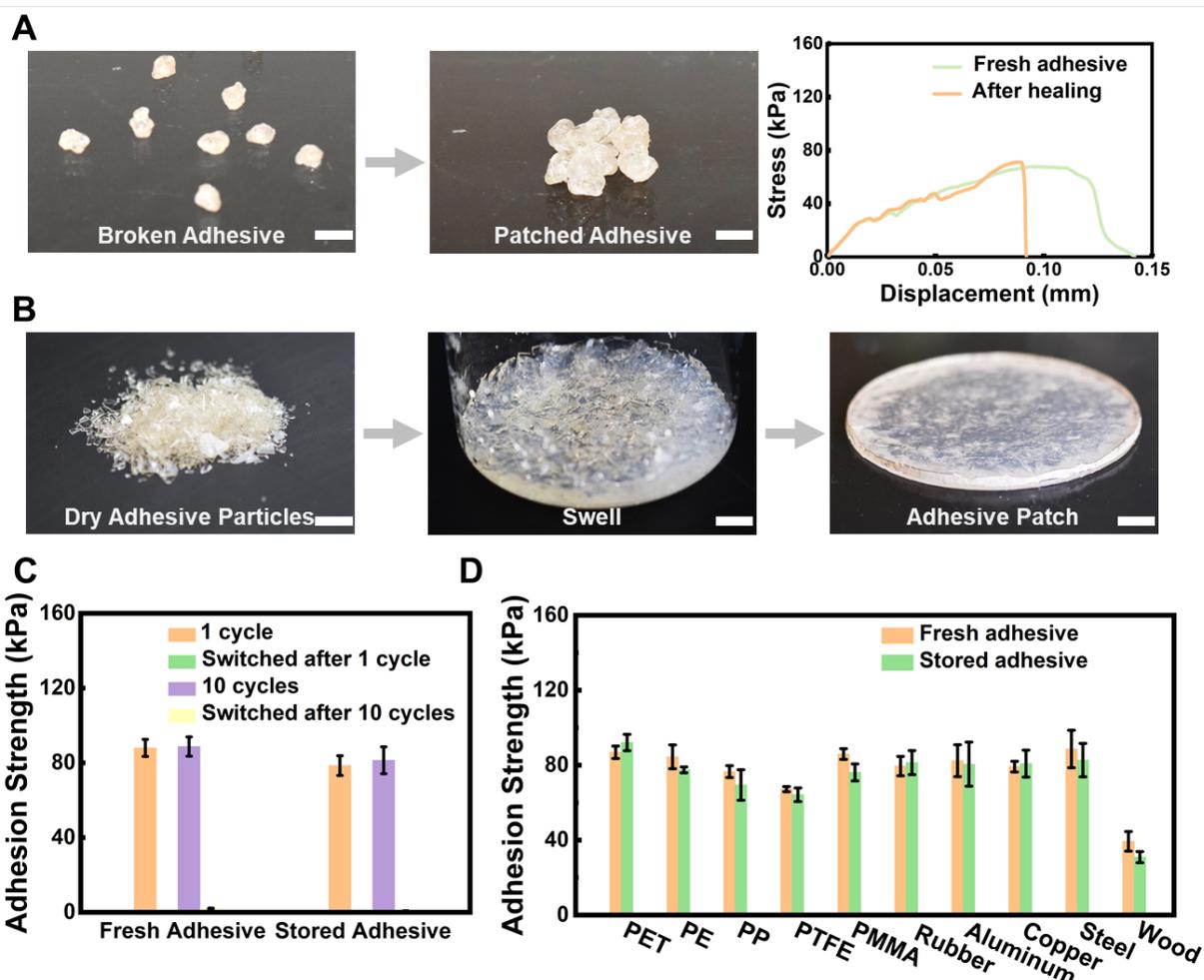

**Figure 4**. Self-healing and dry-stored reusable adhesive. **(A)** Broken adhesive pieces are patched together, reshaped, and allowed to heal for 24 hours. The adhesion strength after healing is nearly identical to that of the freshly made adhesive. **(B)** The adhesive can be dried, grinded into small particles, dry-stored for 10 days, re-swollen in water, patched together, and reused. **(C)** The stored adhesive almost completely recovers its adhesion strength at room temperature and excellent switching at high temperature, even after 10 switching cycles. **(D)** The complete reusability extends to adherends made of diverse materials. All error bars represent the standard deviation of more than 3 independently measured samples. The scale bars in (A) and (B) are 1 cm.

**Selective pick-and-release of objects of various materials, shapes, sizes, and weights.** The strong adhesion and near-vanishing residual adhesion after switching enable pick-and-release of objects of various materials, shapes, sizes, and weights, all with switching time around 10 s (**Table S3**). The adherends demonstrated here include grape tomato (8.7 g, switching time of 4 s), lotus seed (1.1 g, 3 s), wooden rod (0.25 g, 11 s), plastic screw (0.35 g, 5 s), copper screw (1.9 g, 10 s), and steel nail (0.32 g, 9 s) (**Figure 5A, Movie S5**). In addition, with a contact area of 3.14 cm$^2$, the adhesive gripper can lift a total weight of 2.4 kg, equivalent to a stress of 76 kPa (**Figure S17, Movie S6**). Under a programmed heating, this pick-and-release is reversible for multiple cycles with the same switching time of about 10 s, as shown by **Figure 5B** and **Movie S7**.

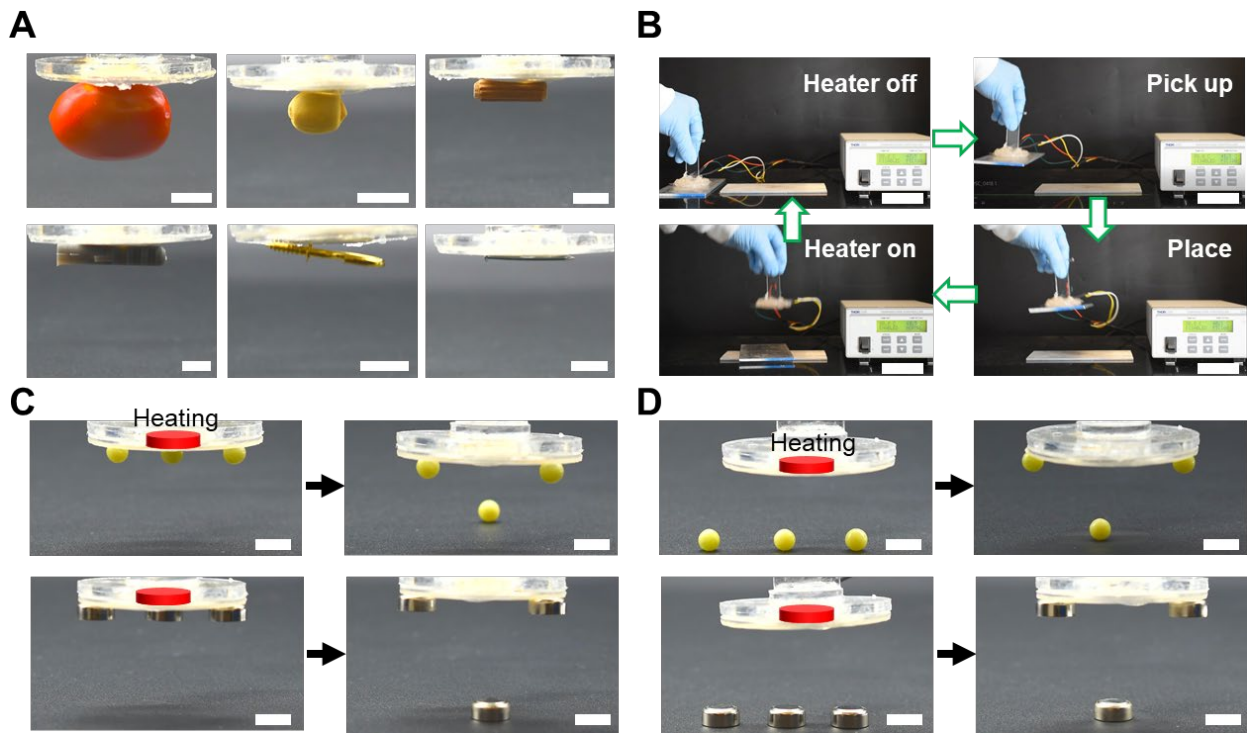

**Figure 5**. Pick-and-release enabled by the strong and switchable adhesive. **(A)** Pick-and-release of objects of various materials, shapes, sizes, and weights, all with switching time around 10 s (**Movie S5**). See **Table S3** for detailed data. The scale bar is 1 cm. **(B)** Under programmed heating, the gripper can pick and release objects for multiple cycles (**Movie S7**). The scale bar is 4 cm. **(C)&(D)** Selective release or pick-up of spherical beads (top) and flat coin batteries (bottom) by only heating the center of adhesive (**Movies S8&S9**). The scale bars are 1 cm.

In showcasing the prospective future applications of the switchable soft sticky adhesive, such as high-throughput assembly manufacturing (4, 5), here we demonstrate a selective pick-and-release mechanism by spatially programming the heating zones on the adhesive gripper. To do so, we place a circular thermal insulator around the central heater in the gripper, such that only the central part of the adhesive layer increases its temperature by heating **(Figure S18)**. When the heater is turned off at room temperature, the adhesive is sticky everywhere, and can pick up all the objects **(Figure 5C left)**. Activating the heater subsequently triggers the selective release of only the central object, with the remaining objects still attached to the gripper **(Figure 5C right & Movie S8)** due to the pre-designed central heating zone. The same setup further enables a selective pick-up when activating the heater during the object attachment, as shown in **Figure 5D and Movie S9**.

**Discussion and Summary**

In summary, we have presented a design approach for strong and highly switchable soft sticky adhesives by synergizing the surface stickiness, viscoplastic bulk energy dissipation, and switchable polymer chain retraction in a thermo-responsive PNIPAm polymer network. The adhesive shows strong adhesion at room temperature and near-vanishing adhesion at elevated temperature with diverse materials. The thermal switching is facile, reversible, and repeatable for many cycles. The switching time in 10 seconds is faster than or comparable to most state-of-the-art switchable adhesives, which can be further reduced by a thinner adhesive layer that enables faster thermal conduction. The adhesive is fully reusable due to its complete self-healing. Integrated with predesigned heating zones on a customized gripper, the adhesive enables versatile pick-and-release of objects with various materials, shapes, sizes, and weights.

In designing a strong or tough irreversible adhesive, the adhesion energy $\Gamma$ is commonly expressed as a sum of two parts, $\Gamma = \Gamma_0[1+f_d(\alpha)]$ (26), where $\Gamma_0$ is the *intrinsic toughness* determined solely by the local process around the interface during debonding, and $f_d$ is an amplifying factor from nonlocal energy dissipation with a state variable $\alpha$ (e.g., loading rate), such as the bulk viscoplastic bond breaking. In the current system, the near-vanishing adhesion at high temperature is a direct result of switching $\Gamma_0$ to nearly zero, through the interfacial chain retraction by the condensed network after phase transition. This classical equation thus provides a new way of designing switchable adhesion with a synergistic division of labor. On the one hand, a strong and tough adhesive can be designed by introducing well-established toughening mechanisms in the bulk that greatly enhance $f_d$, such as the strengthening *and* toughening through interpenetrating polymer networks (e.g., PNIPAm with calcium-alginate (44), poly(acrylic acid) (45), or poly(vinyl alcohol) (46)). Simultaneously, a highly switchable adhesive can be further achieved by eliminating $\Gamma_0$ via stimuli-responsive phase transition, such as the thermo-responsive chain retraction in a broad range of lower critical solution temperature (LCST) (47) and upper critical solution temperature (UCST) polymers (48).

Another future optimization lies in constructing effective surface architectures of these strong and switchable adhesives. A heterogeneous surface architecture, such as a pillar-like structure, can greatly enhance the adhesion due to crack pinning, re-nucleation, and stress de-concentration (49). In addition, a surface architecture can also accelerate the switching due to its small feature size that reduces the time scale of thermal or solvent diffusion (17). To enable such a design that harnesses surface architectures for both adhesion strengthening and switching, the three-way interplay between the architecture geometry, debonding, and adhesion switching calls for further fundamental investigation.

**Materials and Methods**

**Materials for adhesive.** All chemicals were used as received without any further purification. These include *N*-isopropylacrylamide (NIPAm, Tokyo Chemical Industry, I0401), nanoclay (Laponite RD, BYK USA), *N,N,N′,N′*-tetramethylethylenediamin (TEMED, Sigma-Aldrich, T7024), and ammonium persulfate (APS, Sigma-Aldrich, A3678).

**Materials as adherends.** We purchased the following materials as adherends used in the current paper: glass (borosilicate glass sheet, McMaster-Carr), PET (polyethylene terephthalate, McMaster-Carr), PE (polyethylene, McMaster-Carr), PP (polypropylene, Amazon), PTFE (polytetrafluoroethylene, Amazon), PMMA (poly(methyl methacrylate), Amazon), silicone rubber (silicone rubber sheet, 50A durometer, Amazon), aluminum (6061 aluminum sheets, McMaster-Carr), copper (110 copper bars, McMaster-Carr), steel (304 stainless steel, McMaster-Carr), and wood (Amazon).

**Synthesis of adhesives.** For all the PNIPAm adhesives synthesized in this paper, we fix the amount of NIPAm (monomers), and vary the amounts of other chemicals including the nanoclay (crosslinker), APS (initiator), and TEMED (accelerator). First, 6.3 g NIPAm and a certain amount of nanoclay are dissolved into 30 mL deionized water. The mixture is stirred overnight in a sealed container to form a homogeneous and transparent solution. APS and TEMED are further added into the solution and subsequently mixed by a speed mixer (FleckTeck DAC 330) with 2000 rpm for 2 minutes. The precursor solution is then poured into a glass mold and covered with a glass plate for curing at room temperature overnight to complete the polymerization. We investigate the following mass ratios of nanoclay, APS, and TEMED: $M_{clay}/M_{NIPAm}$ = 0.079, 0.158, 0.238, 0.317 and 0.476; $M_{APS}/M_{NIPAm}$ = $1.909 \times 10^{-3}$, $2.863 \times 10^{-3}$, $3.818 \times 10^{-3}$, $5.727 \times 10^{-3}$, and $7.635 \times 10^{-3}$; and $M_{TEMED}/M_{NIPAm}$ = $7.251 \times 10^{-3}$ and $72.51 \times 10^{-3}$. To investigate the effect of water content on adhesion, we synthesize identical PNIPAm gels with 6.3 g NIPAm, 30 mL deionized water, $M_{clay}/M_{NIPAm}$ = 0.238, $M_{APS}/M_{NIPAm}$ = $2.863 \times 10^{-3}$, and $M_{TEMED}/M_{NIPAm}$ = $7.251 \times 10^{-3}$. This corresponds to a water content of $M_{water}/M_{gel}$ = 0.7936 After the synthesis, we use solvent exchange to obtain gels with various water contents. For $M_{water}/M_{gel}$ = 0.912 or 0.930, a gel is swollen in the deionized water. For $M_{water}/M_{gel}$ = 0.310, 0.481, or 0.654, a gel is dried in the fume hood. In each case, the swelling or drying is continued until the gel reaches the desired water content measured by its weight. Afterwards, the gel is stored in a sealed sample bag for 48 hours to reach thermodynamic equilibrium. The weights of all gels are measured again before their final characterization. After exploring the synthesis-property relationship in **Figure 2**, we fix the final composition of the PNIPAm adhesive as $M_{clay}/M_{NIPAm}$ = 0.476, $M_{APS}/M_{NIPAm}$ = $3.818 \times 10^{-3}$, and $M_{TEMED}/M_{NIPAm}$ = $72.51 \times 10^{-3}$, and $M_{water}/M_{gel}$ = 0.7936, for the rest of the paper.

**Characterization of adhesion strength.** The adhesion strength is characterized by the probe-pull experiment (26). In such a test, the PNIPAm adhesive of circular shape with a diameter of 2 cm and thickness of 1 mm thickness is fixed to a circular acrylic sheet on one side by super glue (Loctite), and attached to an adherend on the other side. The sandwiched structure is compressed by a weight of 3 kg (about 95 kPa compressive stress) for 5 minutes to form the adhesion. Afterwards, we mount the structure into an Instron tensile tester (34TM-5), with its top acrylic sheet fixed to a homemade gripper and the bottom adherend to another gripper via the adhesive tape VHB (4905, 3M) (**Figure S2A**). The Instron conducts a tensile test with a fixed speed of 0.1 mm/s to measure the force-displacement curve (**Figure S2B**). The adhesion strength is calculated as the peak force divided by the initial cross-sectional area of the adhesive. The adhesion strengths measured with or without the homemade gripper are nearly identical, as confirmed by our control experiments.

**Characterization of adhesion energy.** A bilayer of PNIPAm adhesive and adherend is formed by direct attachment with the same prescribed compressive force as in the characterization of adhesion strength. The other side of the PNIPAm adhesive is attached with a stiff backing layer made of PET via super glue (Loctite), to constrain the elastic deformation of the adhesive during peeling. The entire composite forms a strip of 5 cm in length and 1 cm in width, with a 0.5 cm interfacial pre-crack between the adhesive and adherend at its one end. The strip is subsequently mounted into the Instron tester (34TM-5) and subjected to a double-peeling test (for flexible adherends including PET, PP, PTFE, and silicone rubber, as illustrated in **Figure S3A**) or a 90-degree peeling test (for stiff adherends including PE, PMMA, aluminum, copper, steel, and wood). In either test, the crack speed is prescribed as 1 mm/s, directly controlled by the peeling speed. The raw data of a measured force-displacement curve is represented in **Figure S3B**. The curve reaches a steady state with a plateau when the peeling displacement increases. The adhesion energy is calculated as $\Gamma = 2F/w$ in double peeling, and $\Gamma = F/w$ for 90-degree peeling (26), where $F$ is the measured steady-state force and $w = 1$ cm is the sample width.

**Thermal imaging.** For the thermal imaging in **Figure S9**, a thermal camera (HIKMICRO Thermal Imaging Camera Pocket 2, 256×192 IR resolution) is used to show the temperature distribution.

**Adhesive gripper with programmed heating.** A homemade adhesive gripper is assembled by laser-cut acrylic parts as shown in **Figure S12**. The gripper is further integrated with a ceramic heater (HT19R Thorlabs) and an *in situ* temperature sensor (TH100PT Thorlabs) on the heater. The heater is programmed by a temperature controller (THORLABS TC 200). The PNIPAm adhesive is attached to the bottom of the heater. During a programmed heating, a prescribed temperature is set by the controller. When the heating switch is turned off on the controller, the adhesive will naturally cool down to room temperature and become sticky again.

**Characterization of switching time.** The top of the homemade adhesive gripper is connected to the Instron gripper while the PNIPAm adhesive at the bottom is attached to a steel dead weight of 250 g with a diameter of 1.7 cm, resulting in a measured stress of about 11 kPa by Instron (**Figure 2C and Figure S13**). In an experiment with thermo-induced switching, the heater is set with a target temperature of 40 °C, and the gripper can hold the dead weight for about 14 s, followed by the sudden drop of the dead weight and thus the measured stress (**Figure S13B** and **Figure 2D**). In another control experiment, the heater is turned off, and the gripper can hold the weight for more than 100 s at room temperature (**Figure S13A**).

**Characterization of cyclic switching.** The cyclic switching of adhesion is characterized following the experimental setup in **Figure S10**. The top of the adhesive gripper is connected to the Instron gripper while the PNIPAm adhesive at the bottom is attached to an adherend. The adherend is fixed to a hot plate, and the hot plate is fixed to the bottom of the Instron tensile tester. In one switching cycle, the PNIPAm adhesive is first brought into contact with the adherend with a compression at room temperature to form the adhesion. The adherend is subsequently heated to 75 °C set by the hot plate. When the adhesive becomes completely opaque due to the thermo-induced phase transition, it is detached from the adherend. A subsequent switching cycle is started after approximately 1 minute to ensure all materials are completely cooled back to room temperature. To measure the adhesion strength and switching performance after a specified number of switching cycles, we prepare multiple identical samples and measure their adhesion strength at room temperature and at high temperature after 1, 3, 5, 7, and 10 switching cycles. These measurements end up taking 30 samples in total (for each specific cycle number, we characterize 3 samples at room temperature and 3 samples at high temperature, to include the statistics). For more switching cycles (i.e., 30 cycles shown in **Figure S11**), a slight amount of deionized water is sprayed to the surface of adhesive after each cycle to keep the adhesive from drying due to the continuous heating.

**Preparation of reused adhesives.** After a probe-pull experiment (**Figure S16**), all the broken adhesive pieces are re-collected and patched together. The patched adhesive is reshaped by a glass mold, compressed with a heavy weight, and allowed to heal for 24 hours. The adhesion strength is measured again after the healing.

**Preparation and reuse of dry-stored adhesives.** Fresh adhesives are dried in a fume hood for 6 hours. The dried adhesives are then grinded into small particles and stored in sealed bags for 10 days. To reuse the dry-stored adhesives, 9.3 g of dry adhesive particles are mixed into 30 mL of water and swell until equilibrium. The swollen adhesive pieces are subsequently patched together, transferred into a glass mold, compressed with a heavy weight, and allowed to heal for 24 hours.

**Characterization of contact angle.** Contact angles of various adherends are measured by a Phoenix 150 contact angle measurement system. In each measurement, deionized water is dripped onto an adherend substrate three times to obtain one data point.

**Acknowledgments**

This work was supported by the Haythornthwaite Research Initiation Grant through the Applied Mechanics Division of American Society of Mechanical Engineers (ASME). R.B. and Q.Y. are grateful to Qihan Liu for insightful discussions on osmocapillary adhesion.

# Supporting Information

# Strong and Highly Switchable Soft Sticky Adhesives

Qianfeng Yin[1], Yilmaz Arin Manav[2], Yichen Wan[1], Benyamin Davaji[2,*], and Ruobing Bai[1,*]

[1]Department of Mechanical and Industrial Engineering, College of Engineering, Northeastern University, Boston, MA, 02115, USA.

[2]Department of Electrical and Computer Engineering, College of Engineering, Northeastern University, Boston, MA, 02115, USA.

*Corresponding author: Benyamin Davaji: b.davaji@northeastern.edu; Ruobing Bai: ru.bai@northeastern.edu

**This PDF file includes:**

Figures S1 to S18

Tables S1 to S3

Reference

**Captions for Movies S1 to S8**

**Movie S1.** Debonding of PNIPAm adhesives on adherends at room temperature and elevated temperature.

**Movie S2.** Process of thermo-induced phase transition in a bulk PNIPAm gel. The white color represents the condensed polymer network at temperature above 32 °C.

**Movie S3.** The adhesive gripper can successfully pick up an aluminum plate at room temperature, but cannot pick up an identical plate that has been preheated to precisely 32 °C.

**Movie S4.** When the temperature of the aluminum plate increases to slightly above 32 °C, the adhesive undergoes phase transition indicated by the color change, and is easily detached from the plate.

**Movie S5.** The adhesive can pick and release objects of various materials, shapes, sizes, and weights.

**Movie S6.** The adhesive can lift a weight of 2.4 kg with a contact area of 3.14 cm$^2$.

**Movie S7.** Pick-and-release of aluminum plates for multiple cycles with a switching time of about 10 s.

**Movie S8.** Selective release of small objects with curved and flat surfaces.

**Movie S9.** Selectively pick-up of small objects with curved and flat surfaces.

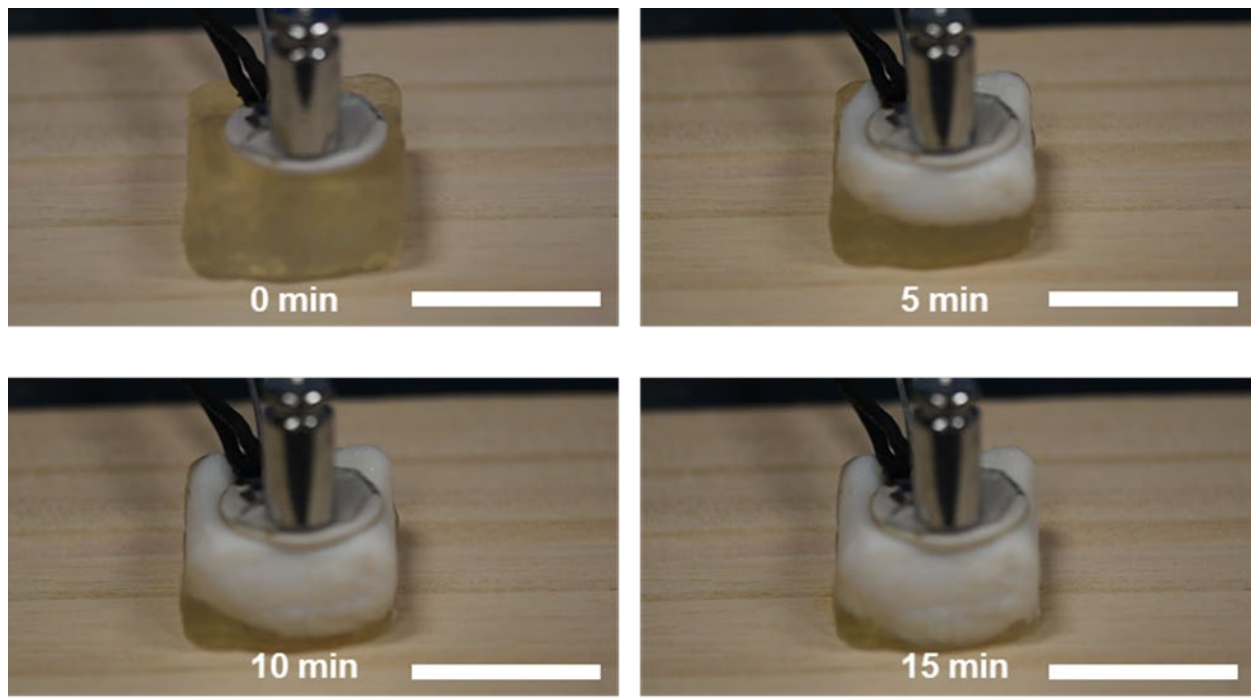

**Figure S1.** Process of thermo-induced phase transition in a bulk PNIPAm gel. The white color represents the condensed polymer network. The heater with temperature set as 90 °C is attached on top of a bulk PNIPAm gel with a thickness of 1.5 cm. The adhesive is translucent initially, and gradually becomes white, indicating the thermo-induced phase transition of the PNIPAm network. The scale bar is 3 cm.

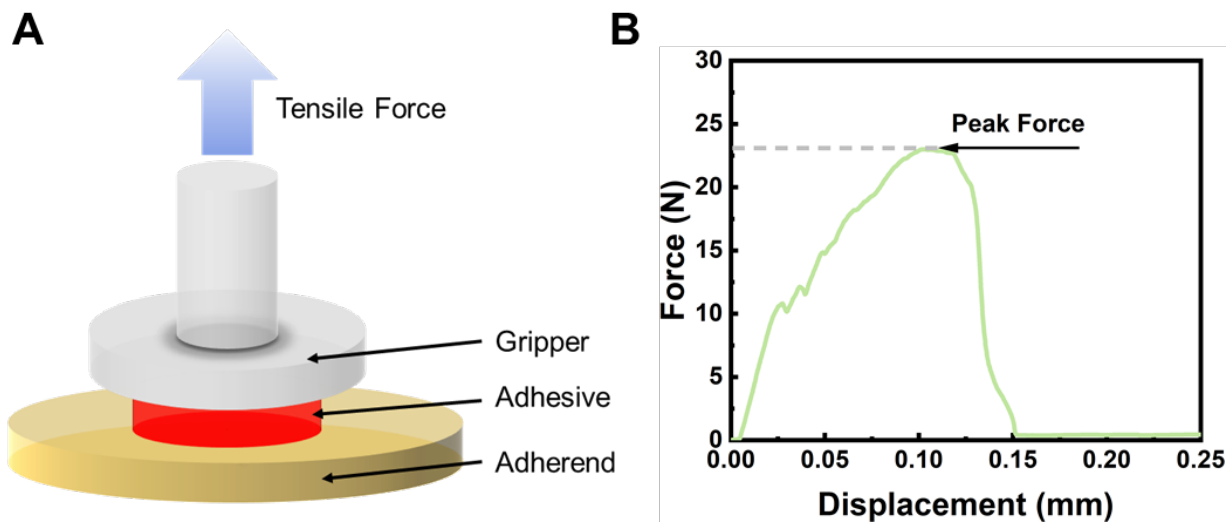

**Figure S2.** Characterization of adhesion strength by the probe-pull experiment (1). **(A)** The homemade gripper and the adherend are fixed to the top and bottom grippers of an Instron tensile tester (34TM-5). The Instron tester conducts a tensile test with a fixed speed of 0.1 mm/s to measure force-displacement curve. **(B)** The measured force-displacement shows a peak force. The adhesion strength is calculated as the peak force divided by the initial cross-sectional area of the adhesive.

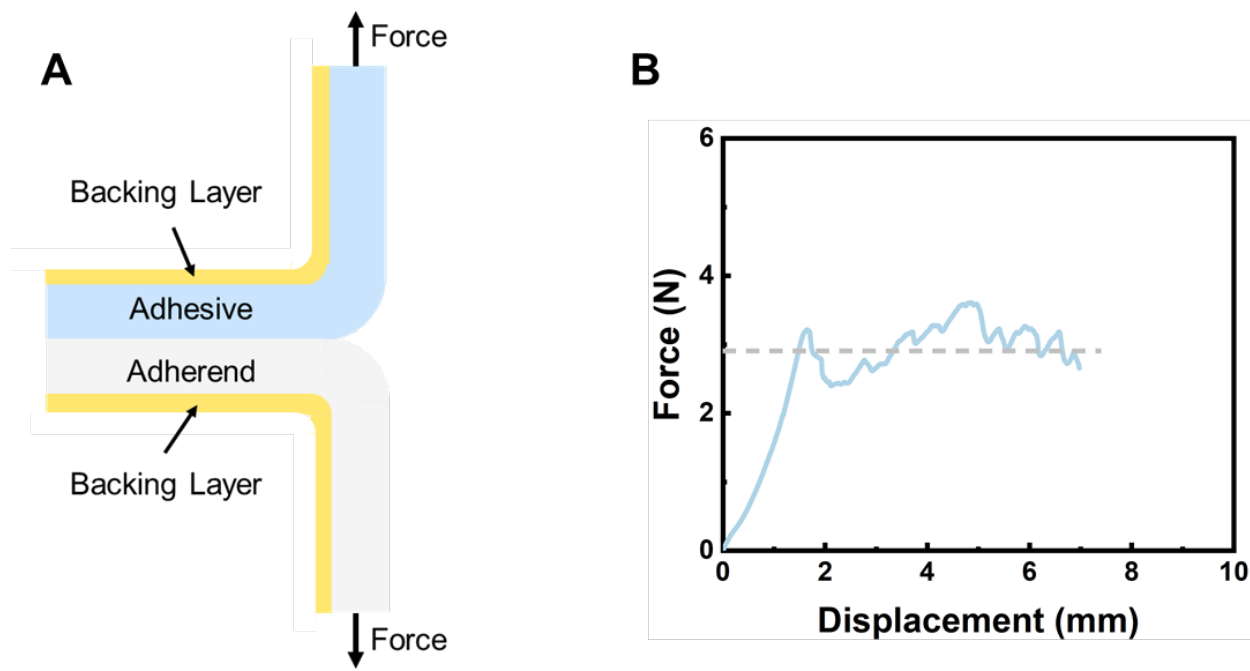

**Figure S3.** Characterization of adhesion energy. **(A)** Schematics of a double-peeling test. The bilayer of adhesive and adherend is attached with stiff backing layers via super glue to constrain the elastic deformation during peeling. For stiff adherends, a 90-degree peeling test is used instead of double-peeling. **(B)** A representative force-displacement curve measured by the peeling test. The crack speed is prescribed as 1 mm/s. The curve reaches a steady state with a plateau when the peeling displacement increases. The adhesion energy is calculated as $\Gamma = 2F/w$ in double peeling, and $\Gamma = F/w$ for 90-degree peeling (1), where $F$ is the measured steady-state force and $w = 1$ cm is the sample width.

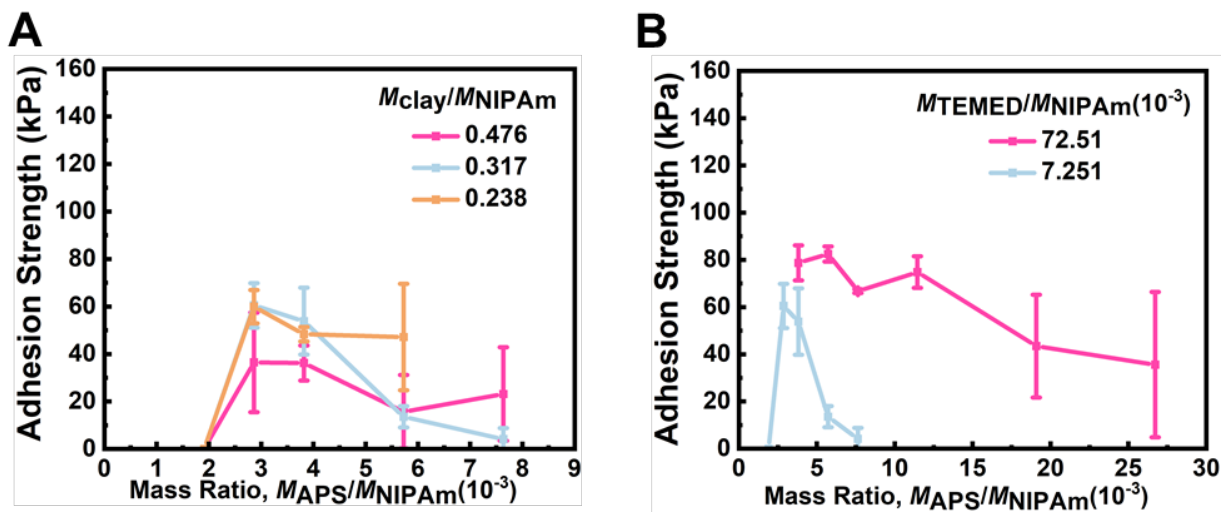

**Figure S4.** The effect of initiator amount, $M_{APS}/M_{NIPAm}$, on the adhesion strength, with different fixed values of **(A)** $M_{clay}/M_{NIPAm}$, with $M_{TEMED}/M_{NIPAm} = 7.251 \times 10^{-3}$, and **(B)** $M_{TEMED}/M_{NIPAm}$, with $M_{clay}/M_{NIPAm} = 0.317$. The adherend is chosen as glass for all the data here. All error bars represent the standard deviation of more than 3 independently measured samples.

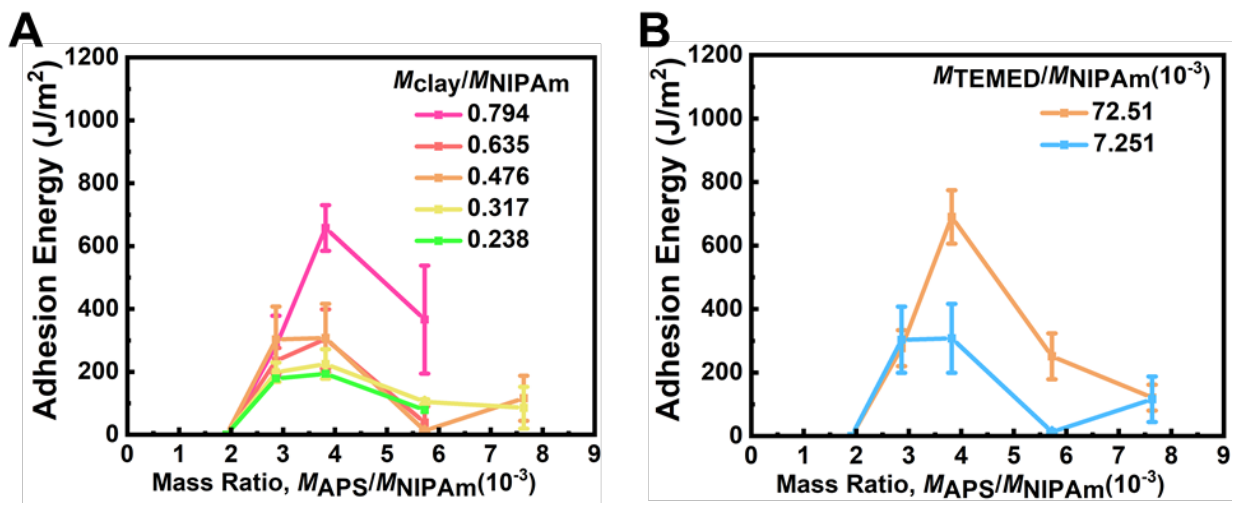

**Figure S5.** The effect of initiator amount, $M_{APS}/M_{NIPAm}$, on the adhesion energy, with different fixed values of **(A)** $M_{clay}/M_{NIPAm}$, with $M_{TEMED}/M_{NIPAm} = 7.251 \times 10^{-3}$, and **(B)** $M_{TEMED}/M_{NIPAm}$, with $M_{clay}/M_{NIPAm} = 0.476$. The adherend is chosen as PET for all the data here. All error bars represent the standard deviation of more than 3 independently measured samples.

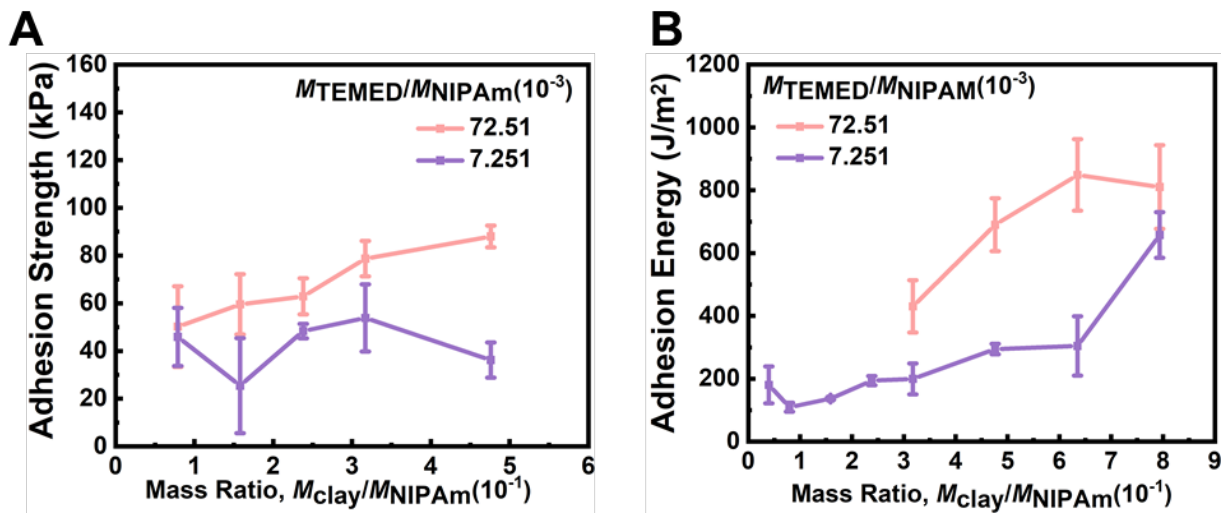

**Figure S6.** The effect of nanoclay amount, $M_{clay}/M_{NIPAm}$, on the **(A)** adhesion strength with glass and **(B)** adhesion energy with PET. The amount of initiator is fixed as $M_{APS}/M_{NIPAm} = 3.818 \times 10^{-3}$. All error bars represent the standard deviation of more than 3 independently measured samples.

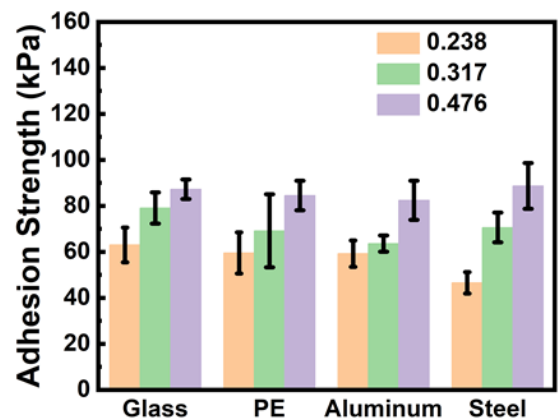

**Figure S7.** Adhesion strength with different adherends and various mass ratios of nanoclay, $M_{clay}/M_{NIPAm}$ = 0.238, 0.317, and 0.476. The amounts of initiator and accelerator are fixed as $M_{APS}/M_{NIPAm}$ = 3.818×10$^{-3}$ and $M_{TEMED}/M_{NIPAm}$ = 72.51×10$^{-3}$, respectively. All error bars represent the standard deviation of more than 3 independently measured samples.

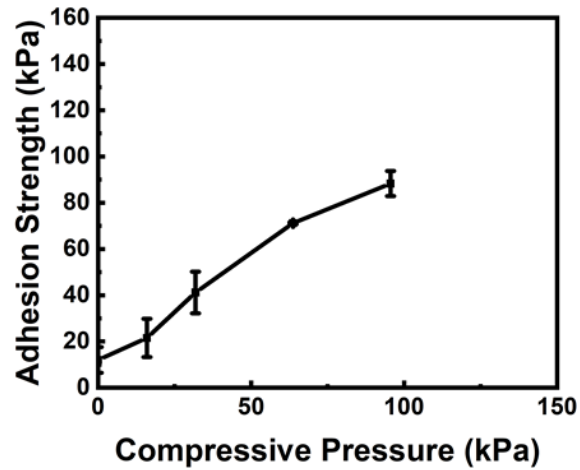

**Figure S8.** The measured adhesion strength increases with the compressive pressure applied to attach the adhesive to a glass adherend. In experiments, various weights (0 kg, 0.5 kg, 1 kg, 2 kg, and 3 kg) are placed on top of the adhesive (circular shape with diameter of 2 cm) for 5 min to provide the compressive pressure. All error bars represent the standard deviation of more than 3 independently measured samples.

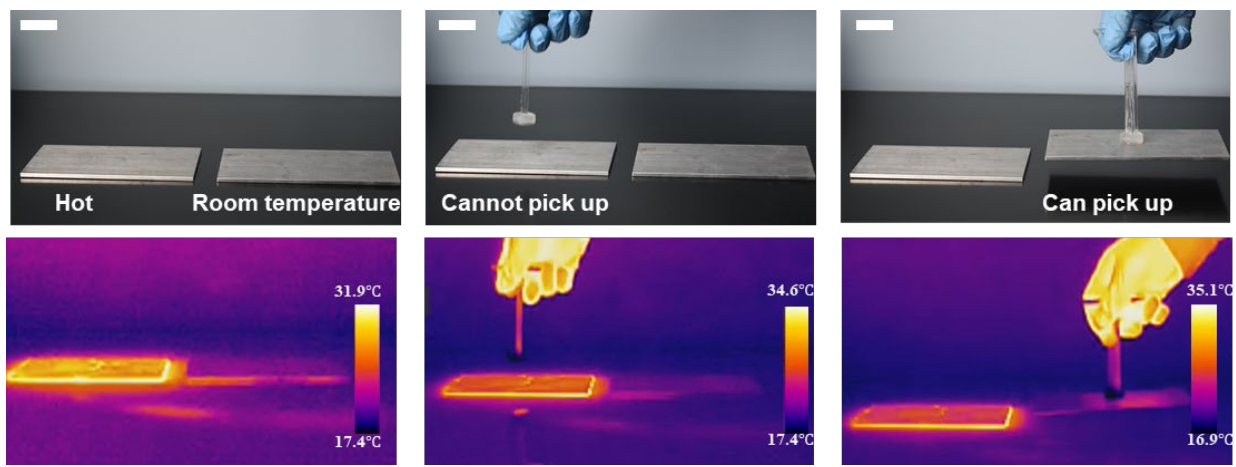

**Figure S9.** The gripper can successfully pick up an aluminum plate at room temperature, but cannot pick up an identical plate that has been preheated to precisely 32 °C. The scale bar is 3 cm.

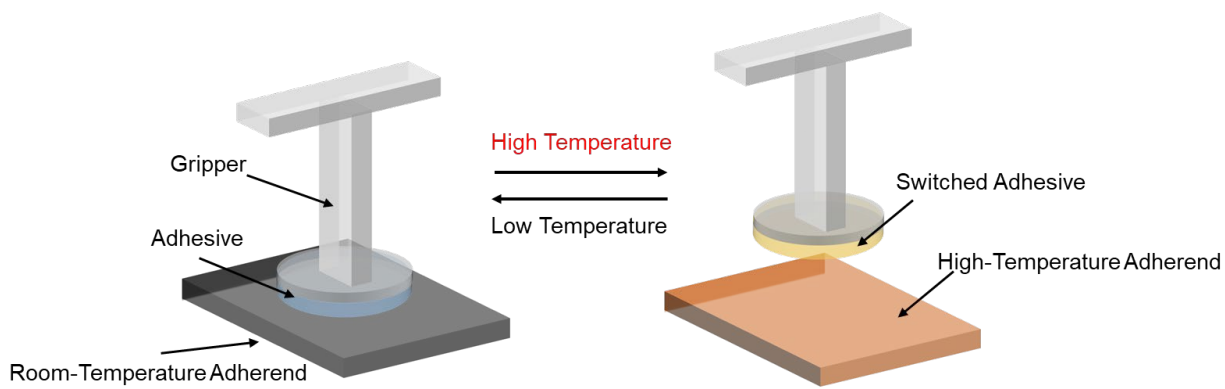

**Figure S10.** Characterization of cyclic switching. The top of the adhesive gripper is connected to the Instron gripper while the PNIPAm adhesive at the bottom is attached to an adherend. The adherend is fixed to a hot plate (not shown in the figure), and the hot plate is fixed to the bottom of the Instron tensile tester. In one switching cycle, the PNIPAm adhesive is first brought into contact with the adherend with a compression at room temperature to form the adhesion. The adherend is subsequently heated to 75 °C set by the hot plate. When the adhesive becomes completely opaque due to the thermo-induced phase transition, it is detached from the adherend. A subsequent switching cycle is started after approximately 1 minute to ensure all materials are cooled back to room temperature.

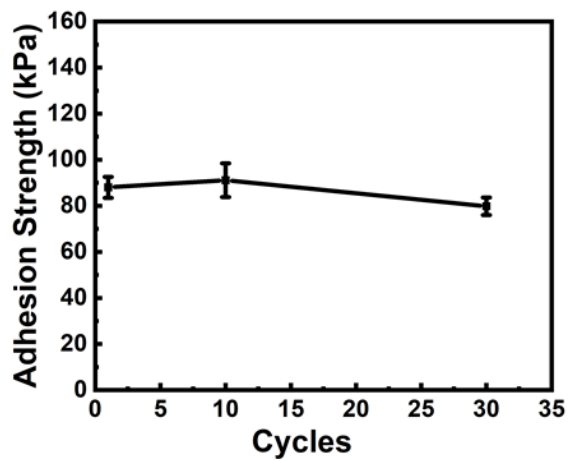

**Figure S11.** The measured adhesion strength over an extended number of switching cycles. A slight amount of deionized water is sprayed to the surface of adhesive after each cycle to keep the adhesive from drying due to the continuous heating. The adhesion strength is measured after 1, 10, and 30 switching cycles. All error bars represent the standard deviation of more than 3 independently measured samples.

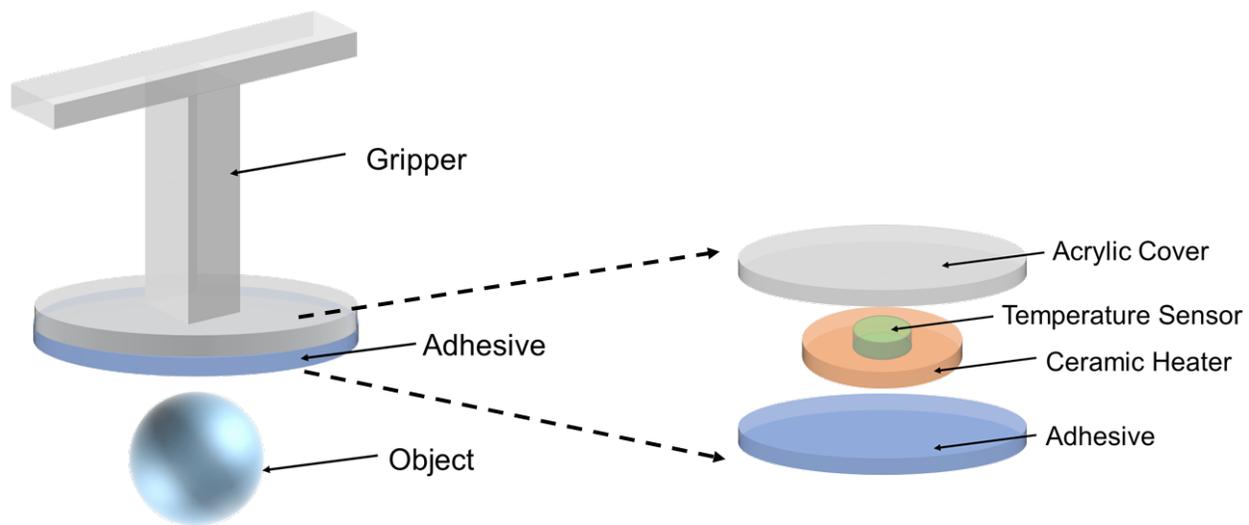

**Figure S12.** Illustration of the homemade adhesive gripper integrated with a ceramic heater and an *in situ* temperature sensor. The heater is programmed by a temperature controller (not shown here). The PNIPAm adhesive is attached to the bottom of the heater.

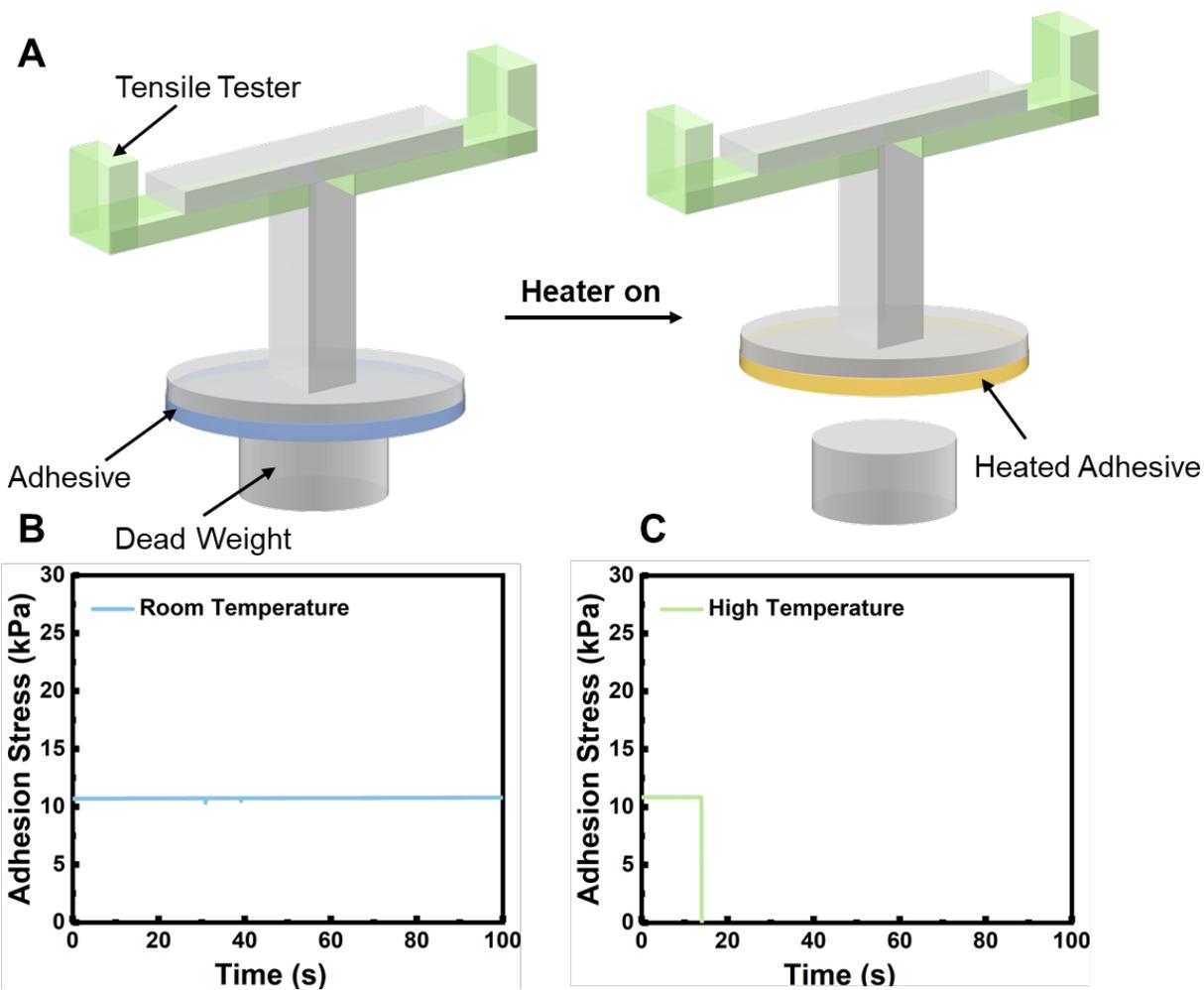

**Figure S13.** Characterization of switching time. **(A)** The top of the homemade adhesive gripper is connected to the Instron tensile tester while the PNIPAm adhesive at the bottom is attached to a steel dead weight of 250 g with a diameter 1.7 cm. When the heater is set with a target temperature of 40 °C, the gripper can hold the dead weight for about 14 s, followed by the drop of the dead weight and thus the measured stress. **(B)** At room temperature, the gripper can hold the weight for more than 100 s, shown by the constant stress plateau of about 11 kPa measured by the tensile tester. **(C)** When the heater is on with a target temperature of 40 °C, the adhesion stress undergoes an abrupt drop after 14 s, which represents the switching time.

**Table S1. Adhesion strength at room and high temperatures after 1 and 10 switching cycles.**

| Adherend | Adhesion Strength (kPa) | | | |
|---|---|---|---|---|
| | 1 cycle | Switched 1 cycle | 10 cycles | Switched 10 cycle |
| Glass | 87.19±4.27 | 0.58±0.16 | 88.72±5.16 | 1.89±0.32 |
| PET | 86.92±3.32 | 1.22±0.18 | 115.77±30.13 | 0.82±0.43 |
| PE | 84.49±6.39 | 0.36±0.08 | 91.57±7.50 | 0.75±0.31 |
| PP | 76.57±3.17 | 1.01±0.60 | 62.55±5.09 | 0.34±0.05 |
| PTFE | 67.24±1.35 | 0.74±0.54 | 60.74±5.10 | 0.75±0.19 |
| PMMA | 85.96±2.82 | 1.33±0.43 | 89.51±12.74 | 0.77±0.37 |
| Silicone Rubber | 79.52±5.15 | 1.130±0.65 | 49.00±15.17 | 0.85±0.24 |
| Aluminum | 82.39±8.50 | 0.91±0.40 | 98.49±3.96 | 0.39±0.27 |
| Copper | 79.27±2.87 | 0.73±0.28 | 78.15±9.52 | 0.66±0.37 |
| Steel | 88.68±9.97 | 0.53±0.33 | 89.11±18.36 | 0.26±0.08 |
| Wood | 39.34±5.22 | 2.82±0.77 | 29.15±5.23 | 0.36±0.04 |

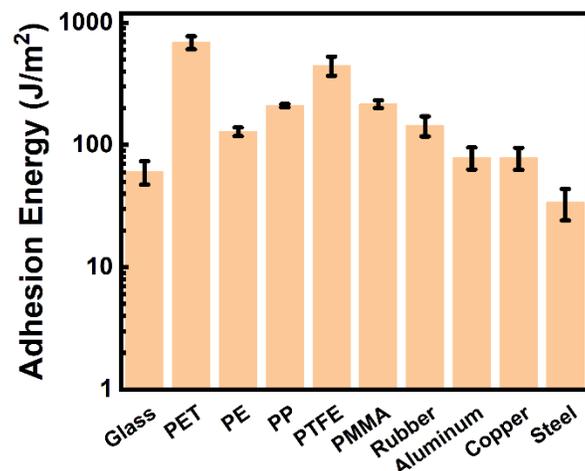

**Figure S14.** Adhesion energy with different adherends. The composition of the PNIPAm adhesive is $M_{Clay}/M_{NIPAm} = 0.476$, $M_{APS}/M_{NIPAm} = 3.818\times10^{-3}$, $M_{TEMED}/M_{NIPAm} = 72.51\times10^{-3}$, and $M_{water}/M_{gel} = 0.7963$. All error bars represent the standard deviation of more than 3 independently measured samples.

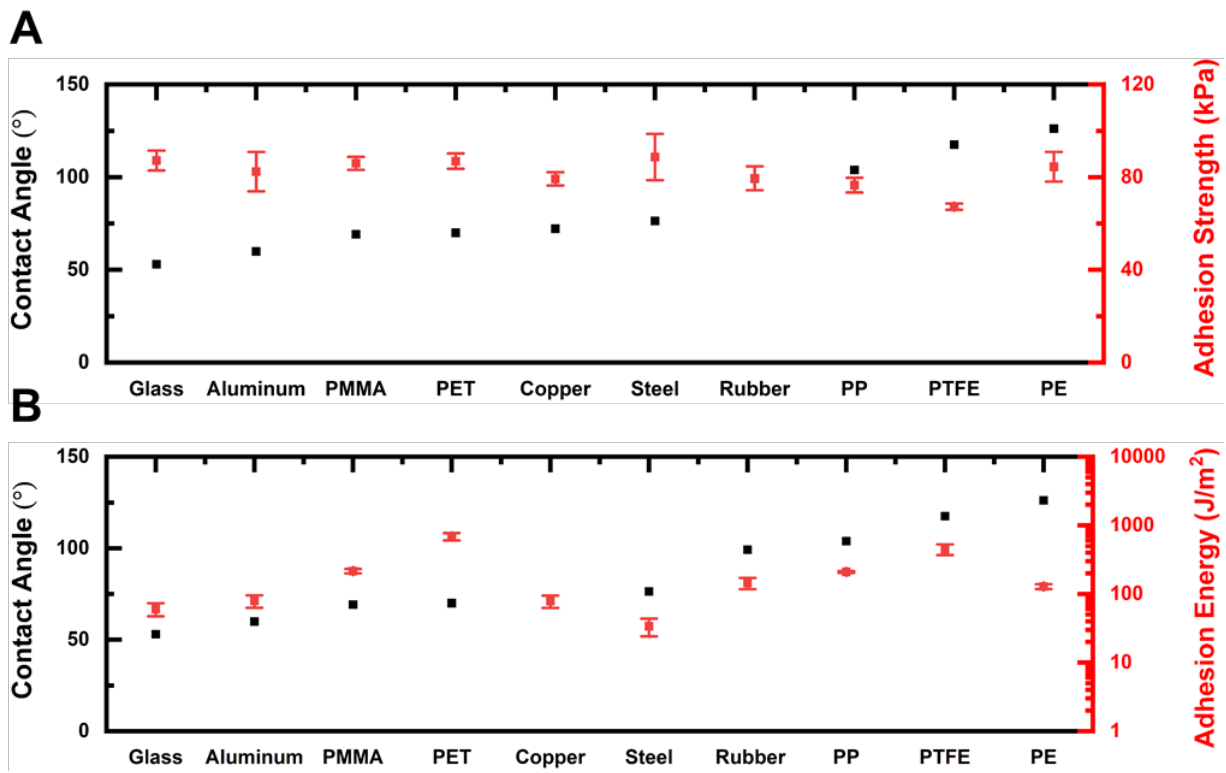

**Figure S15.** Contact angle of various adherends and their corresponding **(A)** adhesion strengths and **(B)** adhesion energies. All error bars represent the standard deviation of more than 3 independently measured samples.

**Table S2. Contact angle, adhesion strength, and adhesion energy of various adherends.**

| Adherend | Contact Angle (º) | Adhesion Strength (kPa) | Adhesion Energy (J/m$^2$) |
|---|---|---|---|
| Glass | 52.95±3.71 | 87.19±4.27 | 60.42±13.05 |
| Aluminum | 59.92±6.61 | 82.39±8.50 | 79.11±16.31 |
| PMMA | 69.17±4.85 | 85.96±2.82 | 216.05±15.89 |
| PET | 69.98±1.77 | 86.92±3.32 | 690.13±84.22 |
| Copper | 72.19±3.30 | 79.27±2.87 | 78.64±15.89 |
| Steel | 76.41±3.56 | 88.68±9.97 | 33.93±9.81 |
| Silicone Rubber | 99.22±2.74 | 79.52±5.14 | 144.51±27.15 |
| PP | 103.87±1.18 | 76.57±3.17 | 209.91±6.11 |
| PTFE | 117.55±1.76 | 67.24 1.35 | 448.84±79.53 |
| PE | 126.24±0.21 | 84.49±6.39 | 209.91±6.11 |

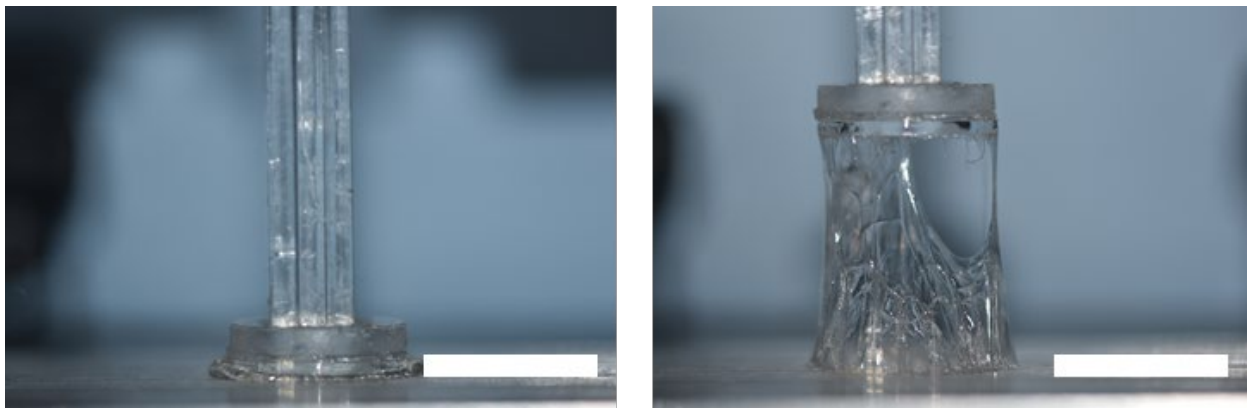

**Figure S16.** Broken adhesive after a probe-pull experiment. The scale bar is 2 cm.

**Table S3. Weight, geometry, and switching time of various objects.**

| Object | Weight (g) | Diameter (mm) | Length (cm) | Switching Time (s) |
|---|---|---|---|---|
| Tomato | 8.7 | 20 | 2.9 | 4 |
| Lotus Seed | 1.066 | 10 | 1.1 | 3 |
| Wood Rod | 0.25 | 5 | 1.9 | 11 |
| Plastic Screw | 0.35 | 6 | 3.8 | 5 |
| Copper Screw | 1.89 | 4 | 2.6 | 10 |
| Steel Nail | 0.32 | 1 | 2.5 | 9 |
| Plastic Bead | 0.37 | 6 | 0.6 | 15 |
| Coin Battery | 1.925 | 10 | 1 | 12 |

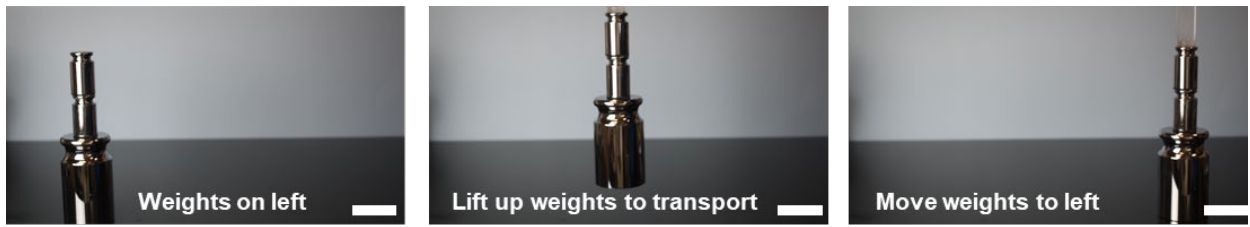

**Figure S17.** With a contact area of 3.14 cm$^2$, the adhesive gripper can pick and release a total weight of 2.4 kg, equivalent to a stress of 76 kPa. The scale bar is 5 cm.

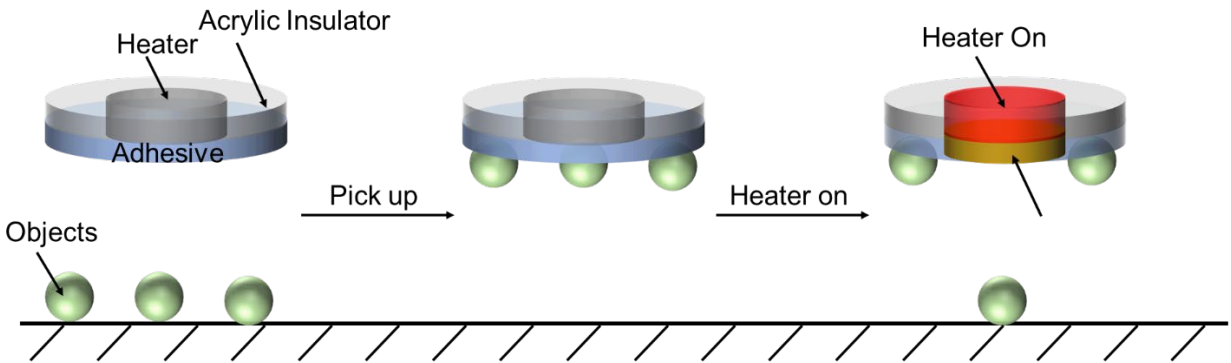

**Figure S18.** Schematics illustrating selective release of small objects. A circular thermal insulator made of acrylic is placed around the central heater in the gripper, such that only the central part of the adhesive layer increases its temperature by heating. When the heater is off at room temperature, the adhesive is sticky everywhere, and can pick up all the objects. When heater is on, only the central heating zone increases its temperature above 32 °C, thus releasing the object inside the zone.